\documentstyle[epsf, rotate]{WORKSHOP}

\newcommand{\XMM}{{\em XMM-Newton }}
\newcommand{\Ch}{{\em Chandra }}
\newcommand{\Su}{{\em Suzaku }}

\def\gappeq{\mathrel{ \rlap{\raise.5ex\hbox{$>$}}
                      {\lower.5ex\hbox{$\sim$}}  } }
\def\lappeq{\mathrel{ \rlap{\raise.5ex\hbox{$<$}}
                      {\lower.5ex\hbox{$\sim$}}  } }

\begin{document}

\title{
{\it Suzaku} Observations of the Radio Galaxy 3C\,33 \\ 
}

\author{
Daniel A. Evans$^1$, James N. Reeves$^2$, Martin J. Hardcastle$^3$,\\ 
Ralph P. Kraft$^4$, Julia C. Lee$^4$, and Shanil N. Virani$^5$
\\[12pt]  
%
$^1$  MIT Kavli Institute for Astrophysics and Space Research, 77 Massachusetts Avenue, Cambridge, MA 02139, USA \\
$^2$  Astrophysics Group, School of Physical and Geographical Sciences, Keele University, Keele, ST5 5BG, UK \\
$^3$  School of Physics, Astronomy \& Mathematics, University of Hertfordshire, College Lane, Hatfield AL10 9AB, UK \\
$^4$  Harvard-Smithsonian Center for Astrophysics, 60 Garden Street, Cambridge, MA 02138, USA \\
$^5$  Department of Astronomy, Yale University, P.O. Box 208101, New Haven, CT 06520, USA \\
{\it E-mail (DE): devans@space.mit.edu} 
}

\abst{
We present results from a new 100-ks \Su observation of the nearby radio galaxy 3C\,33, and investigate the nature of absorption, reflection, and jet production in this source. We model the 2--70 keV nuclear continuum with a power law that is absorbed either through one or more layers of pc-scale neutral material, or through a modestly ionized pc-scale obscurer. The expected signatures of reflection from a neutral accretion disk are absent in 3C\,33: there is no evidence of a relativistically blurred Fe K$\alpha$ emission line, and no Compton reflection hump above 10~keV. We discuss the implications of this for the nature of jet production in 3C\,33.
}

\kword{workshop: proceedings -- galaxies: active -- galaxies: jets -- galaxies: individual (3C 33) -- X-rays: galaxies}
\maketitle
\thispagestyle{empty}

\section{Overview}

The origin of jets in active galactic nuclei (AGN) is one of the most important unsolved problems in extragalactic astrophysics. While 90\% of all AGN (Seyfert galaxies and radio-quiet quasars) show little or no jet emission, the remaining 10\% (the radio-loud AGN and radio-loud quasars) launch powerful, relativistic twin jets of particles from their cores. Since jets transport a significant fraction of the mass-energy liberated during the accretion process, sometimes out to $\sim$Mpc distances, understanding how they are produced is key to a complete picture of accretion and feedback in AGN.

X-ray observations of the nuclei of radio-loud and radio-quiet AGN are essential for establishing the connection between the accretion flow and jet. Continuum X-ray observations of radio-loud AGN have mostly been restricted to bright broad-line radio galaxies (BLRGs) and quasars, which are oriented relatively close to the line of sight with respect to the observer. In these sources, unabsorbed non-thermal emission from the jet could potentially contaminate the unabsorbed accretion-related X-ray spectrum and thus dilute the apparent strength of the Compton reflection component. {\it Narrow}-line radio galaxies (NLRGs), such as 3C\,33, on the other hand, which are oriented at low to intermediate angles, have the distinct advantage that (unabsorbed) jet-related X-ray emission can be readily spectrally separated from (heavily absorbed) accretion-related emission, allowing a direct measurement of the strength of Compton reflection. However, narrow-line radio galaxies tend to be relatively faint X-ray sources compared to broad-line objects. \Su observations of these sources, therefore, are particularly useful, owing to the high effective area of the X-ray Imaging Spectrometer (XIS) and Hard X-ray Detector (HXD) instruments.

Here, we present the results from a 100-ks \Su observation of one of the brightest NLRGs, 3C\,33, the only such source known so far to show potential evidence for Compton reflection in its 2--10 keV X-ray spectrum. Previous observations of 3C\,33 with \Ch and \XMM (Kraft et al. 2007) showed that its continuum spectrum could not be adequately modeled by the combination of a heavily obscured power law and a soft power law normally fitted to narrow-line FRII sources, due to the large residuals present between 2--4 keV. They suggested that one possibility is that the source is reflection-dominated, and that a Compton-reflection continuum is present in the 2--10 keV spectrum. However, the limited energy range of {\it Chandra} and {\it XMM-Newton} meant that the evidence for reflection in 3C\,33 could be unambiguously claimed.

\section{Suzaku Observations}

We observed 3C\,33 with \Su on 2007 December 26 (OBSID 702059010) for a nominal exposure of 100~ks. Both the X-ray Imaging Spectrometer (XIS) and Hard X-ray Detector (HXD) were operated in their normal modes. The source was positioned at the nominal aimpoint of the HXD instrument. The data were processed using v. 2.1.16 of the \Su processing pipeline, which includes the latest Charge Transfer Inefficiency (CTI) correction applied for the XIS. We used the standard cleaned events files. We extracted the spectrum of 3C\,33 from the XIS CCDs using a source-centered circle of radius 2.5$'$, with background sampled from an adjacent region free from any unrelated sources, as well as the $^{55}$Fe calibration sources at the corners of each detector. For the HXD, we added the NXB and CXB spectra together using the {\sc mathpha} tool. Using the appropriate response file for epoch 4 of \Su observations, {\rm ae\_hxd\_pinhxnome4\_20080129.rsp}, we binned the source spectrum to 3$\sigma$ above the total (NXB + CXB) background level.

\begin{figure}[t]
\centering
\epsfxsize=10cm
\epsfbox{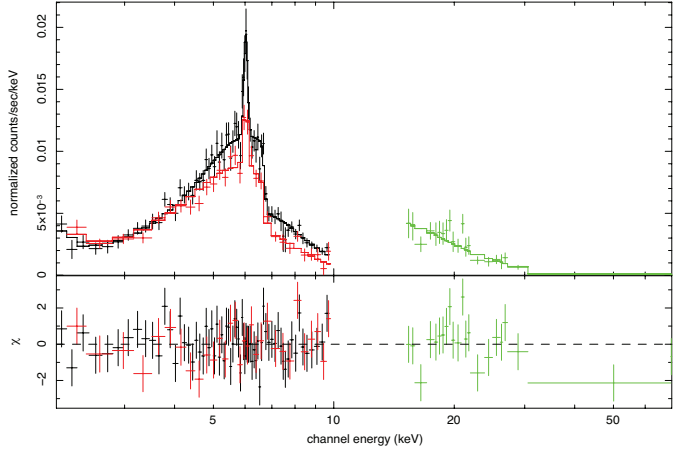}
\caption{ \Su XIS FI {\it (black)}, XIS BI {\it (red)}, and HXD PIN {\it (green)} spectra and residuals of 3C\,33. The model fit is double partially covered primary law of photon index $\Gamma=1.77$, and a Gaussian Fe~K$\alpha$ emission line. }
\label{3pl_3pc}
\end{figure}

\section{Suzaku Spectral Analysis}

We restricted the energy range for our spectral fits to 2--10 keV (XIS) and 15--70 keV (PIN). We fitted to 3C\,33 our canonical model for the X-ray spectrum of NLRGs: the combination of a heavily absorbed power law (likely to be associated with the accretion flow) and a soft, unabsorbed power law (Evans et al. 2006). The unabsorbed power law is simply a parametrization and we defer the discussion of its origin until Section~\ref{softband}. This model gave a relatively poor fit to the spectrum: $\chi^2=150$ for 116 dof, with clear residuals at $\sim$3~keV. The photon index of the primary power law $\Gamma=1.45$ is atypically flat with respect to similar radio galaxies studied by Evans et al. (2006), and the strong residuals at energies $\sim$3~keV likely necessitate a more complex spectral model. We discuss such models next.

\subsection{Model I: Complex Neutral Absorption}

We investigated the possibility that an {\it additional} layer of cold absorption is required to fit the broadband spectrum of 3C\,33 adequately. We adopted a double partial-covering model, implemented in {\sc XSPEC/ISIS} as \textrm{phabs$\times$zpcfabs(1)$\times$zpcfabs(2)$\times$(powerlaw+zgauss)}. We included a Gaussian to represent the neutral Fe~K$\alpha$ line.This model provides a good fit to the spectrum ($\chi^2$=112 for 114 dof). The best-fitting parameters of this model are a power law of photon index $\Gamma$=$1.77^{+0.19}_{-0.10}$, absorbed by columns $N_{\rm H, 1}$=$(2.1^{+0.6}_{-0.5})\times10^{23}$~cm$^{-2}$ ($f_1$=$91^{+3}_{-4}$\%) and $N_{\rm H, 2}$=$(6.0^{+1.7}_{-1.9})\times10^{23}$~cm$^{-2}$ ($f_2$=$80^{+6}_{-8}$\%). Figure~\ref{3pl_3pc} shows the \Su spectrum and residuals for this model. 

We determined the strength of any Compton reflection in the spectrum of 3C\,33 by replacing the primary power law with a {\sc pexrav} component. This neutral reflection model provided no improvement in the fit ($\chi^2$=112 for 113 dof). The best-fitting value of the reflection fraction is $R=0$, with a 90\%-confidence upper limit of R<0.5.

\begin{figure}[t]
\centering
\epsfxsize=10cm
\epsfbox{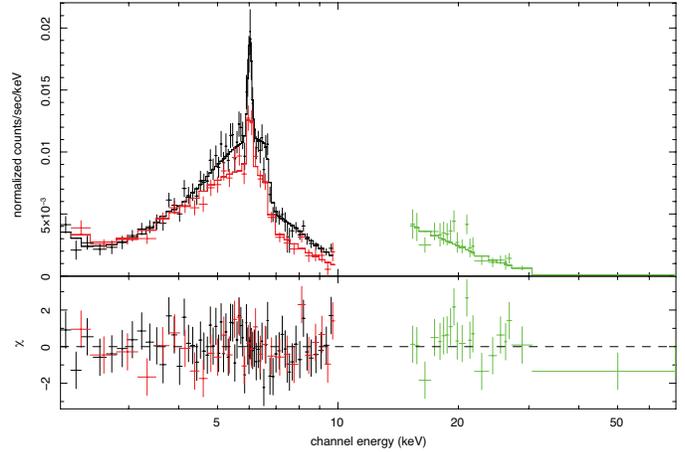}
\caption{ \Su XIS FI {\it (black)}, XIS BI {\it (red)}, and HXD PIN {\it (green)} spectra and residuals of 3C\,33 in the energy range 2--70 keV. The model consists of a power law modified by a single zone of warm absorption, together with a soft, unabsorbed power law, and a Gaussian Fe~K$\alpha$ emission line.  }
\label{xstar}
\end{figure}

\subsection{Model II: Warm Absorption}

An alternative method of modeling the observed spectral complexity in 3C\,33 is to allow the absorber to be partially ionized. We modeled the \Su spectrum as a power law, covered by a single zone of partially photoionized absorption, plus a second, unabsorbed power law to account for the low-energy flux. We included a Gaussian to represent the neutral Fe~K$\alpha$ line. This model provided an excellent fit to the spectrum ($\chi^2=110$ for 115 dof), with a column density $N_{\rm H}$$\sim$7.7$\times10^{23}$~cm$^{-2}$ of modestly ionized gas [log$(\xi)$$\sim$1.1 ergs~cm~s$^{-1}$] as its best-fitting absorption parameters. The best-fitting photon index of the power law is $\Gamma = 2.00\pm0.12$. We show the \Su spectrum, this model, and residuals of $\chi$ in Figure~\ref{xstar}. Again, we determined the strength of any Compton reflection in the spectrum of 3C\,33, and found the best-fitting value of the reflection fraction to be $R=0$, with a 90\%-confidence upper limit of R<0.5.

\begin{figure*}[t]
\centering
\epsfxsize=18cm
\epsfbox{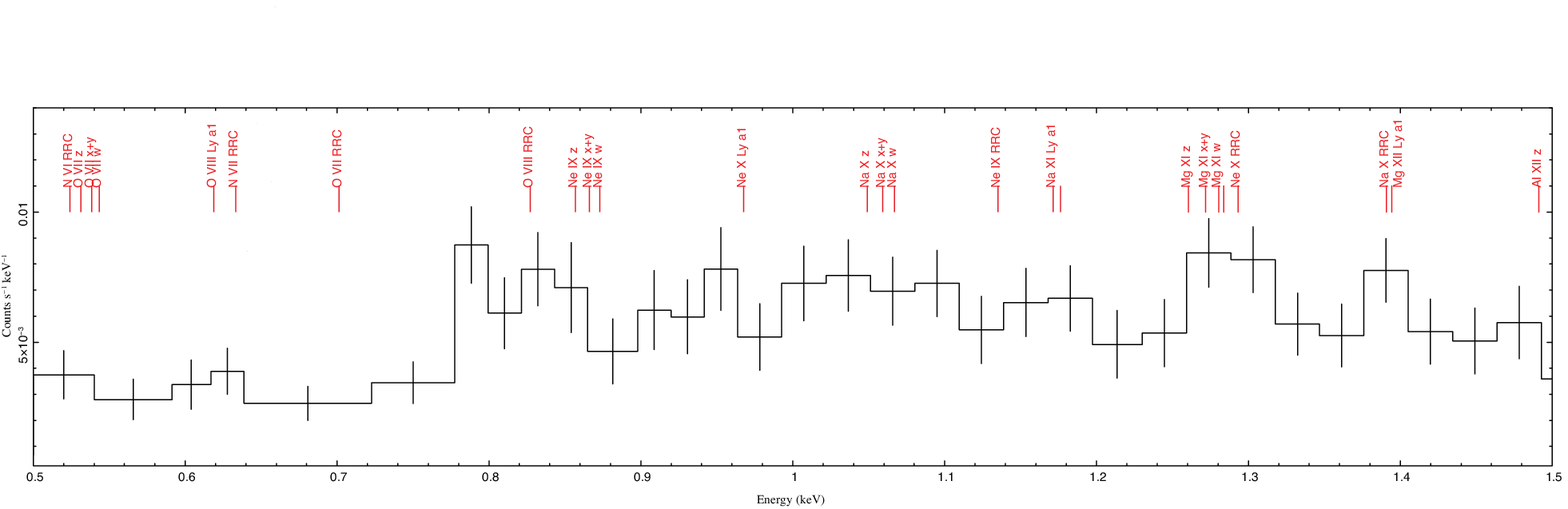}
\caption{ Co-added \Su XIS FI and BI spectrum of 3C\,33 in the energy range 0.5--1.5 keV. The data have been grouped to a minimum signal-to-noise of 3 and to their HWHMs. Overplotted are the lines one would expect to detect in a photoionized plasma, including RRCs, and H- and He-like species. }
\label{photoionized}
\end{figure*}

\subsection{Fe K$\alpha$ Diagnostics}
\label{interp-fek}

The strong Fe~K$\alpha$ line detected with \Su allows us to place constraints on the location and physical state of the fluorescing material. The energy of the line core, $6.385^{+0.021}_{-0.023}$~keV, is consistent with Fe fluorescence from neutral or near-neutral (up to $\sim$Fe~{\sc XVIII}) species. The unresolved Fe K$\alpha$ line width and the estimated black hole mass of $\sim$$4\times10^{8}$ M$_\odot$ (Smith et al. 1990; Bettoni et al. 2003) provide a lower limit to the emission radius of $\gappeq0.1$~pc ($\sim$2000~R$_{\rm s}$), using Keplerian arguments. Finally, the equivalent width of the Fe~K$\alpha$ line is consistent with transmission through an absorbing column of relative iron abundance $A_{\rm Fe}$=1.0 and column density similar to that measured from our spectral fitting (Miyazaki et al. 1996).  There is no evidence for an additional velocity broadened component of the Fe~K$\alpha$ line, which rules out the presence of relativistically blurred fluorescent emission from the innermost portions of the accretion flow. This may imply that a standard, cold Shakura-Sunyaev accretion disk is truncated at a large distance from the central supermassive black hole. We cannot rule out the presence of a modestly broadened ($R\sim100R_{\rm s}$), ionized Fe~K$\alpha$ line in 3C\,33 but nonetheless it seems clear that fluorescent emission from the inner disk is absent.

\section{Soft X-ray Emission: Continuum or Photoionized Gas?}
\label{softband}

There has been considerable debate as to whether the soft X-ray emission in 3C\,33 is dominated by continuum emission from the accretion flow or jet (e.g., Kraft et al. 2007), or rather originates in a photoionized plasma (Torresi et al. 2009). The deep \Su spectrum helps us to resolve this issue. In Figure~\ref{photoionized} we show the co-added \Su FI and BI spectrum from our observation, grouped to a minimum signal-to-noise of 3 and to their HWHMs. We also plot the potential transitions that would indicate that the the soft-band X-ray emission originates in an ionized plasma. For a photoionized plasma, these features include H- and He-like species, as well as narrow Radiative Recombination Continuua (e.g. Evans et al. 2006b). There is tentative evidence for He-like transitions of Ne IX that were first reported by Torresi et al. (2009), and potentially some structure around the He-like triplet of Mg XI. Detailed spectral fitting in this band, however, is necessary to investigate in detail evidence for photoionized gas in the \Su spectrum of 3C\,33.

\section{Interpretation: Compton Reflection in 3C\,33 and other Radio-Loud AGN}

Our results demonstrate that 3C\,33 shows no signs of Compton reflection from neutral material in the inner regions of an accretion disk: there is no reflection hump at energies $>$10~keV, and no evidence for relativistically broadened Fe~K$\alpha$ emission. Indeed, the narrow Fe K$\alpha$ line is likely to originate in a much farther region, at least 2,000~R$_{\rm s}$ from the black hole. 

These characteristic features of inner-disk reflection, which unified models of AGN predict to be prominent, are commonly observed in radio-quiet sources (Seyfert galaxies and radio-quiet quasars) (Reeves et al. 2006; Nandra et al. 2007), but appear to be systematically weaker in radio-loud AGN (e.g., Reeves \& Turner 2000; Evans et al. 2006),  as we see in 3C\,33. 

Several models have been proposed to explain the weakness of Compton reflection continuua in radio-loud AGN, including the idea that accretion disks are highly ionized in high $\dot{M}$ systems (Ballantyne et al. 2002). In this case, the reflection fraction remains high ($R$$\sim$1), but the ionized accretion-disk surface results in a low {\it measured} $R$. An alternative interpretation (Garofalo et al. 2009) is that the black holes in `high-excitation radio galaxies', such as 3C\,33, have retrograde spin with respect to their accretion disks. This pushes the innermost stable circular orbit (ISCO) of the accreting material outwards, meaning that inner-disk reflection is greatly suppressed. Although we cannot distinguish between the two models in the case of 3C\,33, we advertise here that we are currently working on a global interpretation of the cosmological evolution and implications of black-hole spin in radio galaxies (Garofalo, Evans, \& Sambruna 2009).

\section*{References}

\re
Ballantyne D. et al. 2002, MNRAS., 332, 45

\re
Bettoni D. et al. 2003 A\&A., 399, 869

\re
Evans D. A. et al. 2006, ApJ., 642, 96

\re
Evans D. A. et al. 2006b, ApJ., 653, 1121

\re
Garofalo D. et al. 2009, ApJ., 699, 400

\re
Garofalo D., Evans D. A., \& Sambruna R. M., in prep.

\re
Kraft R. P. et al. 2007 ApJ., 659, 1008

\re
Miyazaki S. et al. 1996, PASJ., 48, 801

\re
Nandra K. et al. 2007, MNRAS., 382, 194

\re
Reeves J.~N., \& Turner M.~J.~L. 2000, MNRAS., 316, 234

\re
Reeves J. N. et al. 2006, AN., 327, 1079

\re
Smith E. P. et al. 1990 ApJ., 356, 399

\re
Torresi E. et al. 2009, A\&A., 498, 61

\label{last}

\end{document}